%% file: charm.tex
\documentclass[12pt]{article}
\usepackage[utf8]{inputenc} 
\usepackage[T1]{fontenc}
\usepackage{graphicx}
\include{metadata} 
\usepackage{booktabs}
\usepackage{dcolumn}
\usepackage{float}
\usepackage{amssymb}
\usepackage{array} 
\usepackage{textcomp}
\usepackage{amsmath}
\usepackage{tabularx}
\usepackage{verbatim}
\usepackage{url}
\usepackage{color}
\usepackage{hyperref}
\usepackage{lineno}
\usepackage{multirow}
\usepackage{braket}
\usepackage{caption}
\usepackage{subcaption}
\usepackage{wrapfig}

\newcommand{\babar}{\mbox{\ensuremath{{\displaystyle B}\!{\scriptstyle A}\!{\displaystyle B}\!{\scriptstyle {A\!R}}}\,\,}}
\newcommand{\BDKshh}{$B^- \to D^0 (\to K^0_S \pi^+ \pi^-) \mu^- \bar{\nu}_{\mu}$~}
\newcommand{\BDKshhLL}{$B^- \to D^0 (\to K^0_S  \, \mathrm{(LL)} \, \pi^+ \pi^-) \mu^- \bar{\nu}_{\mu}$~}
\newcommand{\BDKshhDD}{$B^- \to D^0 (\to K^0_S  \, \mathrm{(DD)} \, \pi^+ \pi^-) \mu^- \bar{\nu}_{\mu}$~}
\newcommand{\DKshh}{$D^0 \to K^0_S \pi^+ \pi^-$~}

\newcommand{\DKsPiPiLL}{$D^0 \to K^0_S \, \mathrm{(LL)} \,\pi^+ \pi^-$~}

\def\pbnr{}
\def\speaker{Stefanie Reichert}
\def\onbehalfof{the LHCb collaboration}
\def\title{Time-dependent amplitude analysis of semileptonically-tagged \DKshh decays at LHCb}
\def\affiliation{School of Physics and Astronomy\\
The University of Manchester, Manchester, UK}

\input charmmacros.tex

\begin{document}
\begin{titlepage}
\pubblock

\vfill
\Title{\title}
\vfill
\Author{\speaker\OnBehalf{\onbehalfof}}
\Address{\affiliation}
\vfill
\begin{Abstract}
The hadronic decay $D^0 \to K^0_S \pi^+ \pi^-$ provides access to the measurement of the mixing parameters of the neutral $D$-meson system and allows to test for CP violation. A measurement of the mixing parameters $x_D$ and $y_D$ as well as of the parameters $|q/p|$ and $\phi$, which govern indirect CP violation, will be performed based on a time-dependent amplitude-model analysis of the full LHCb dataset of 2011 and 2012 corresponding to an integrated luminosity of 3 fb$^{-1}$.
\end{Abstract}
\vfill
\begin{Presented}
\venue
\end{Presented}
\vfill
\end{titlepage}
\def\thefootnote{\fnsymbol{footnote}}
\setcounter{footnote}{0}

\section{Introduction}

In the framework of the Standard Model of Particle Physics, the hadronic decay \DKshh provides access to the measurement of the mixing parameters in the neutral $D$ meson system. Due to a cancellation via the GIM mechanism \cite{GIM} and CKM suppression \cite{Cabibbo, CKM}, mixing is suppressed in the neutral charm sector and therefore experimentally challenging. The Standard Model predicts CP violation in $D^0-\bar{D}^0$ mixing (indirect CP violation) to be $\leq 10^{-5}$ \cite{Cicerone}. More recent theoretical calculations \cite{Lenz} find sizable effects on the limit \cite{Cicerone} due to the corrections from the leading Heavy Quark Expansion (HQE) \cite{ManoharWise} contributions. Evidence for New Physics (NP) could be found if discrepancies between experimental observations and the predicted level of indirect CP violation in the Standard Model occur.\\

A measurement of the mixing parameters $x_D$ and $y_D$ as well as of the parameters $|q/p|$ and $\phi = arg(q,p)$, which govern indirect CP violation, will be performed based on a time-dependent amplitude-model analysis of the full LHCb dataset of 2011 and 2012 corresponding to an integrated luminosity of $3 \, \mathrm{fb^{-1}}$. This analysis will combine both prompt and semileptonically-tagged \DKshh decays to obtain a large sample of high purity and with sensitivity at all $D^0$ decay times. Sensitivities of $0.23\%$ for $x_D$, of $0.17\%$ for $y_D$, of $0.2$ for $|q/p|$ and of $11.7^{\circ}$ for $\phi$ are expected for the combined dataset of $3 \, \mathrm{fb^{-1}}$ \cite{Implications}.\\

The current world-averages provided by the {\it{Heavy Flavour Averaging Group}} \cite{HFAG} are $x_D = (0.49^{+0.17}_{-0.18}) \, \%$, $y_D = (0.74 \pm 0.09) \, \%$, $|q/p| = (0.69^{+0.17}_{-0.14})$ and $\phi = (-29.6^{+8.9}_{-7.5}) ^{\circ}$ allowing for CP violation.

\section{Theory}

The hadronic decay $D^0 \to K^0_S \pi^+ \pi^-$ provides access to the measurement of the mixing parameters of the neutral $D$-meson system 

\begin{align}
x_D &= \frac{2(m_1-m_2)}{\Gamma_1 + \Gamma_2}, \\
y_D &= \frac{\Gamma_1 - \Gamma_2} {\Gamma_1 + \Gamma_2}.
\end{align}

since the $D^0$ meson can either decay directly or indirectly after oscillating into a $\bar{D}^0$ affecting the $D^0$ decay time distribution.\\

The linear superposition of the flavour eigenstates $\Ket{D^0}$ and $\Ket{\bar{D}^0}$ yielding the mass eigenstates $\Ket{D_1}$ and $\Ket{D_2}$

\begin{align}
\Ket{D_{1,2}} &=  p \Ket{D^0} \pm q \Ket{\bar{D}^0}
\end{align}

implies indirect CP violation or CP violation in mixing if $|q/p| \neq 1$ while direct CP or CP violation in decay is manifested in a difference in the rates of a decay and its charge-conjugated decay, e.g. $\Gamma(D^0 \to K^0_S \pi^+ \pi^-) \neq \Gamma(\bar{D}^0 \to K^0_S \pi^+ \pi^-)$. In consequence, a time-dependent amplitude analysis is sensitive to the parameters $|q/p|$ and $\phi = arg(q,p)$ governing indirect CP violation whereas a time-integrated analysis of \DKshh decays grants access to direct CP violation.

\section{The LHCb experiment at the Large Hadron Collider}

The analysed data are recorded by the LHCb experiment at the Large Hadron Collider (LHC) \cite{LHC_main, LHC_general, LHC_injector} at CERN. Pure proton beams are produced by stripping off  the electron of hydrogen atoms and the protons are then initially accelerated by the linear collider (LINAC 2). Successively, the protons are accelerated further in the Booster, the Proton Synchrotron (PS) and the Super Proton Synchrotron (SPS) before being injected into the LHC and reaching their final collision energy.\\
 
\begin{figure}[H]
\centering
\includegraphics[width=.85\textwidth]{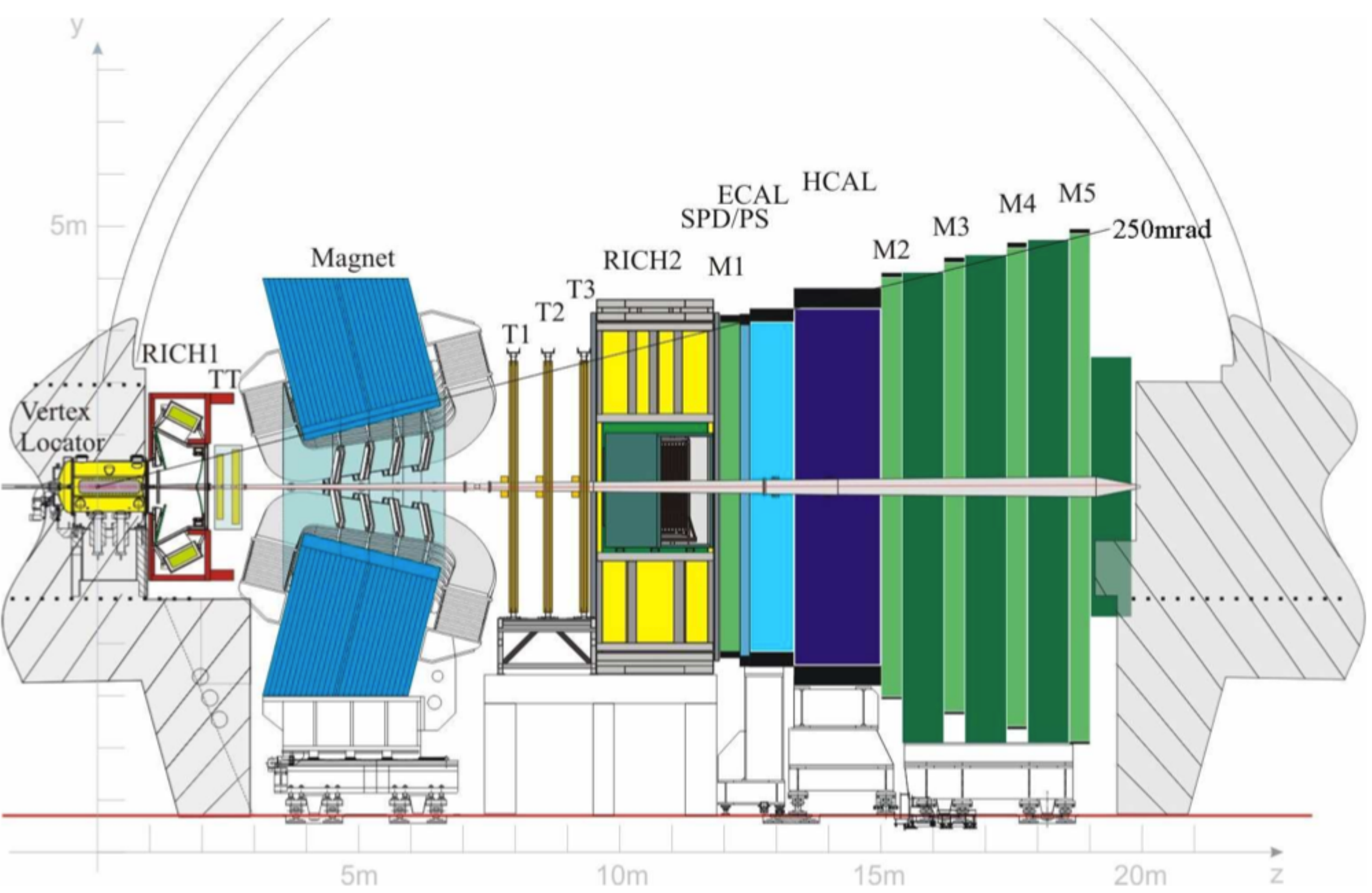}
 \caption[The LHCb detector]{The LHCb detector. The figure is taken from Ref. \cite{LHCb}.}
  \label{Experiment:LHCb}
 \end{figure}

The description of the LHCb detector is based on Ref. \cite{LHCb}. The LHCb detector is a single-arm spectrometer illustrated in Fig. \ref{Experiment:LHCb} where the collision point is chosen as the origin of a right-handed coordinate system depicted in Fig. \ref{Experiment:LHCb}. With a cross section of $10\%$ of all visible events inside the LHCb detector acceptance for producing charm quarks \cite{xsec}, the LHCb detector is perfectly suited for studies in the charm system.\\
The beam pipe is enclosed by the Vertex Locator (VELO) aligned such that the collision point of the protons is located in the centre of the $x$-$y$ plane of the VELO. Built of silicon strip sensors, the VELO provides measurements of track and vertex coordinates with high precision. Apart from the VELO, the tracking system consists of the Tracker Turicensis (TT), and three tracking stations (T1-T3) subdivided into the Inner and Outer Trackers, (IT) and (OT). The TT as well as the IT are composed of silicon microstrip sensors whereas the OT employs straw tubes. Embedded in a $4 \, \mathrm{Tm}$ dipole magnet, this system provides both track coordinate and momentum measurements. Approximately a third of all $K^0_S$ mesons decay inside the VELO acceptance - so-called long tracks (L) - and exhibit a better momentum resolution than so-called downstream tracks (D) of $K^0_S$ mesons decaying outside the VELO acceptance. A system of Ring Imaging Cherenkov Detectors (RICH) is used to obtain excellent separation between kaons and pions. The $\mathrm{C_4F_{10}}$ and aerogel radiators of RICH1 ensure particle identification for low-momentum charged particles whereas the $\mathrm{CF_4}$ radiator of  RICH2 show a better performance for the high-momentum range. The shashlik calorimeter system composed of scintillating tiles and lead absorbers in the Electromagnetic Calorimeter (ECAL) and iron absorbers in the Hadronic Calorimeters (HCAL),  respectively, provide identification and energy measurements of electrons, photons and hadrons. To identify electrons in the trigger, a Scintillator Pad Detector (SPD) and a Preshower Detector (PS) are installed in front of the ECAL. The muon system (M1-M5) uses multi wire proportional chambers filled with a gas mixture of $\mathrm{Ar:C0_2:CF_4}$ and in the inner region of M1 triple-GEM detectors to measure the spatial coordinates of muon tracks and between the M2 to M5 stations, iron blocks serve as absorbers.

\section{Analysis}

At LHCb, \DKshh decays are accessible through various decay channels (charge conjugation is implied throughout). A sample of $D^{* \, +} \to D^0 \pi^+$ decays produced directly in the proton-proton collisions (prompt) has a high yield due to the high production cross section. Due to a cut on the $D^0$ decay time to suppress secondary $D^{* \, +} \to D^0 \pi^+$ decays, only $D^0$ decay times above $0.2 \, \mathrm{ps}$ are accessible. In comparison, the trigger efficiency of semileptonically-tagged \DKshh decays is high for all $D^0$ decay times. A subsample from $\bar{B}^0 \to D^{* \, +} \mu^- \bar{\nu}_{\mu}$ decays is characterised by a cleaner signature due to the additional information from the $D^*$ decay than the sample from $B^- \to D^0 \mu^- \bar{\nu}_{\mu}$ decays. In the following, the discussion will be restricted to the analysis of \DKshh decays originating from $B^- \to D^0 \mu^- \bar{\nu}_{\mu}$ decays.\\

\begin{figure}[H]
\centering
\begin{subfigure}{.45\textwidth}%
\includegraphics[width=1.\textwidth, angle=0]{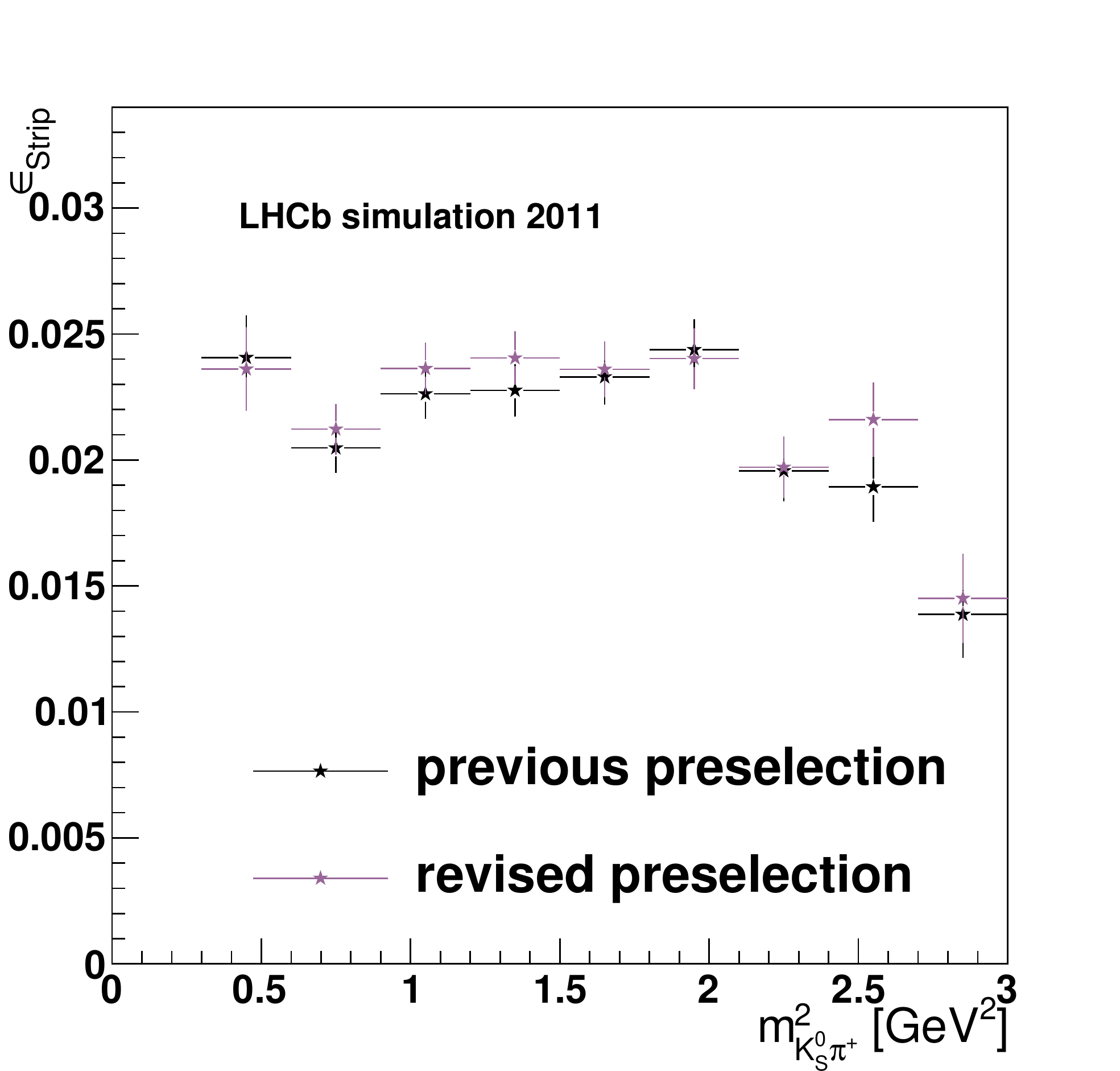}
\captionof{figure}{Versus $m^2_{K^0_S\pi^+}$}
\end{subfigure}%
\begin{subfigure}{.45\textwidth}%
\includegraphics[width=1.\textwidth, angle=0]{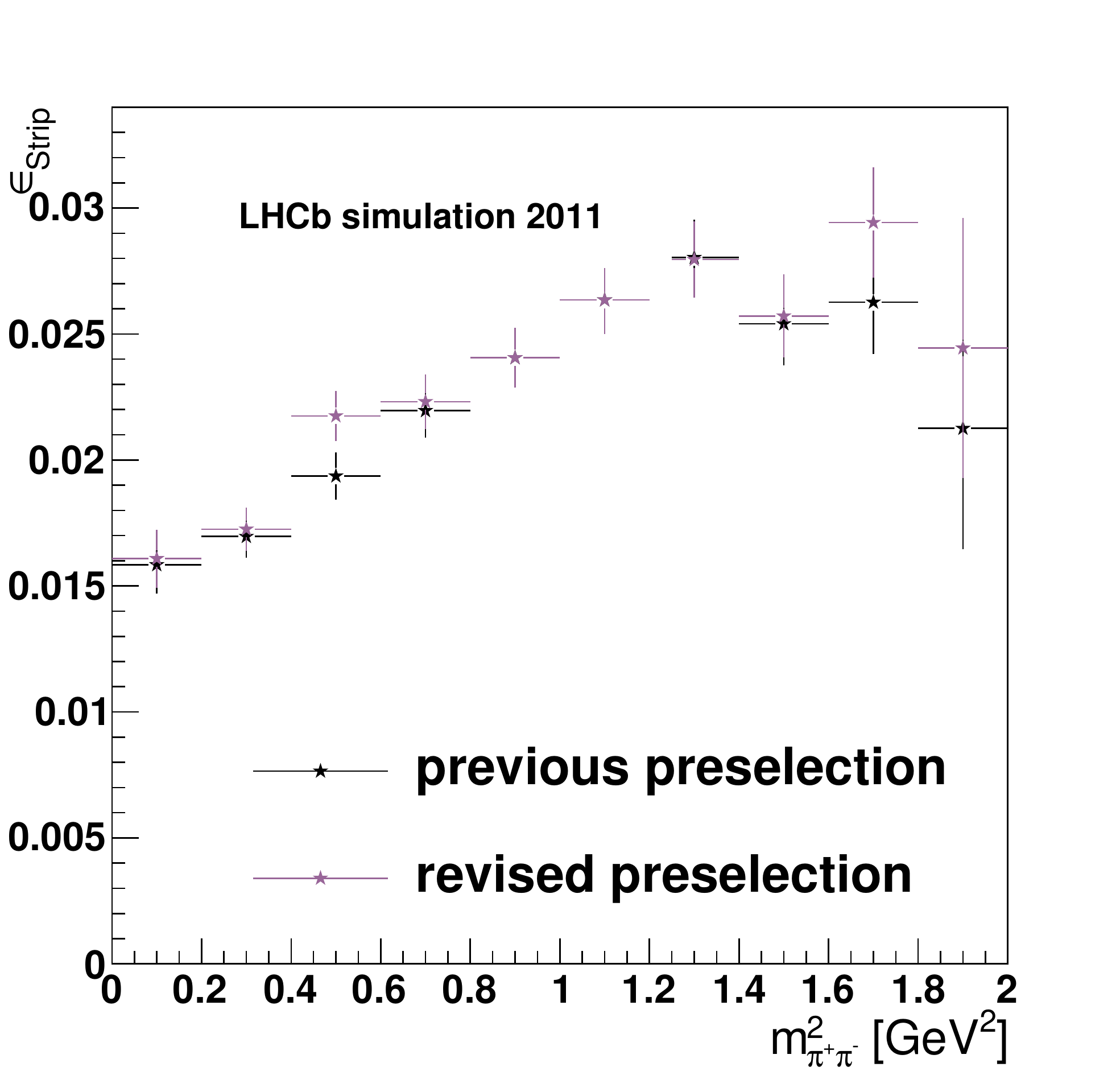}
\captionof{figure}{Versus $m^2_{\pi^+\pi^-}$}
\end{subfigure}%
 \caption{Efficiencies of the previous and revised preselection relative to a phase-space Monte-Carlo simulation sample of \DKsPiPiLL decays originating from $B^- \to D^0 \mu^- \bar{\nu}_{\mu}$ decays studied on a Monte-Carlo simulation sample for 2011 running conditions.}
\label{Stripping_eff}
 \end{figure}

\begin{figure}[H]
\centering
\begin{subfigure}{.45\textwidth}%
\includegraphics[width=1.\textwidth, angle=0]{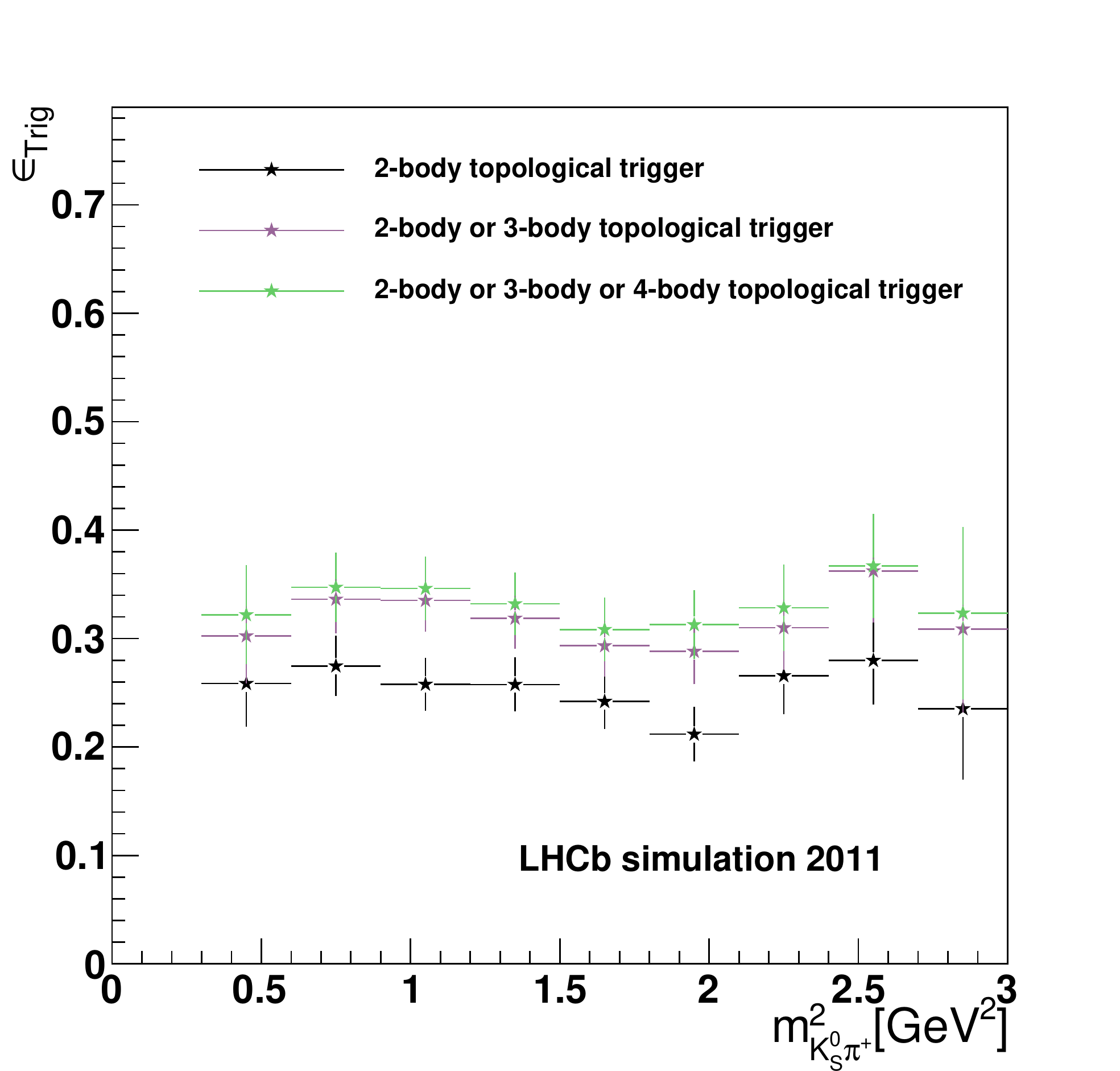}
\captionof{figure}{Versus $m^2_{K^0_S\pi^+}$}
\end{subfigure}%
\begin{subfigure}{.45\textwidth}%
\includegraphics[width=1.\textwidth, angle=0]{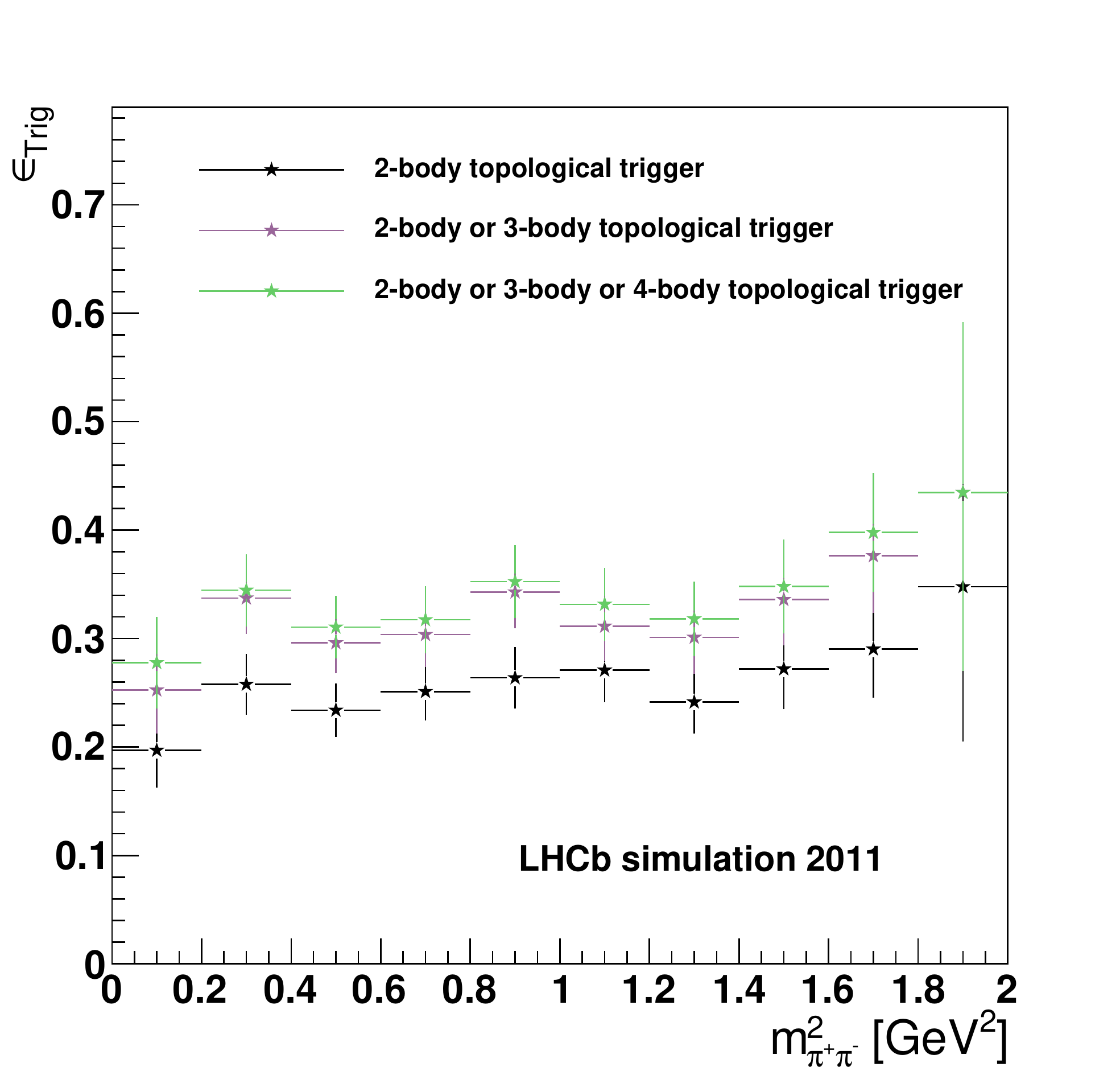}
\captionof{figure}{Versus $m^2_{\pi^+\pi^-}$}
\end{subfigure}%
 \caption{Efficiencies of various topological trigger requirements relative to the \DKsPiPiLL candidates originating from $B^- \to D^0 \mu^- \bar{\nu}_{\mu}$ decays having passed the revised preselection studied on a Monte-Carlo simulation sample for 2011 running conditions.}
 \label{Hlt2_eff}
 \end{figure}

The complete LHCb dataset is passed through an offline preselection algorithm after triggering to select specific decay types. The hardware trigger exploits the transverse momentum of the muon whereas the software-based second stage requires the muon candidate to pass criteria on momentum, transverse momentum, the track $\chi^2/dof$ and impact parameter amongst others. The dataset is then reduced using selection criteria which were revised to increase the efficiency especially at the boundaries of the Dalitz plane and to flatten the efficiency across the Dalitz plane. A uniform efficiency in the Dalitz variables $m^2_{K^0_S\pi^+}$, $m^2_{K^0_S\pi^-}$ and $m^2_{\pi^+\pi^-}$ is required to minimise efficiency corrections and thus also the systematic uncertainty corresponding to the correction. The events selected by the preselection are required to pass through an offline selection and are then used to train a multivariate classifier to further supress background events. A comparison of the previous and revised preselection efficiencies for \BDKshhLL decays and the trigger efficiency for various topological trigger lines measured relative to the \BDKshhLL candidates having passed the preselection are illustrated in Fig. \ref{Stripping_eff} and in Fig. \ref{Hlt2_eff}, respectively.\\

As implied in Fig. \ref{Stripping_eff} and in Fig. \ref{Hlt2_eff}, a slight increase of the efficiency for \BDKshhLL candidates was achieved but variations in the preselection efficiency as a function of $m^2_{K^0_S\pi^+}$ and $m^2_{\pi^+\pi^-}$, respectively, are still observed. In case of the \BDKshhDD, the variations introduced by the trigger lines are large and thus a variation in efficiency is inevitable leading to sizeable efficiency corrections.

 \begin{wrapfigure}{r}{0.45\textwidth}
\centering
\includegraphics[width=.45\textwidth]{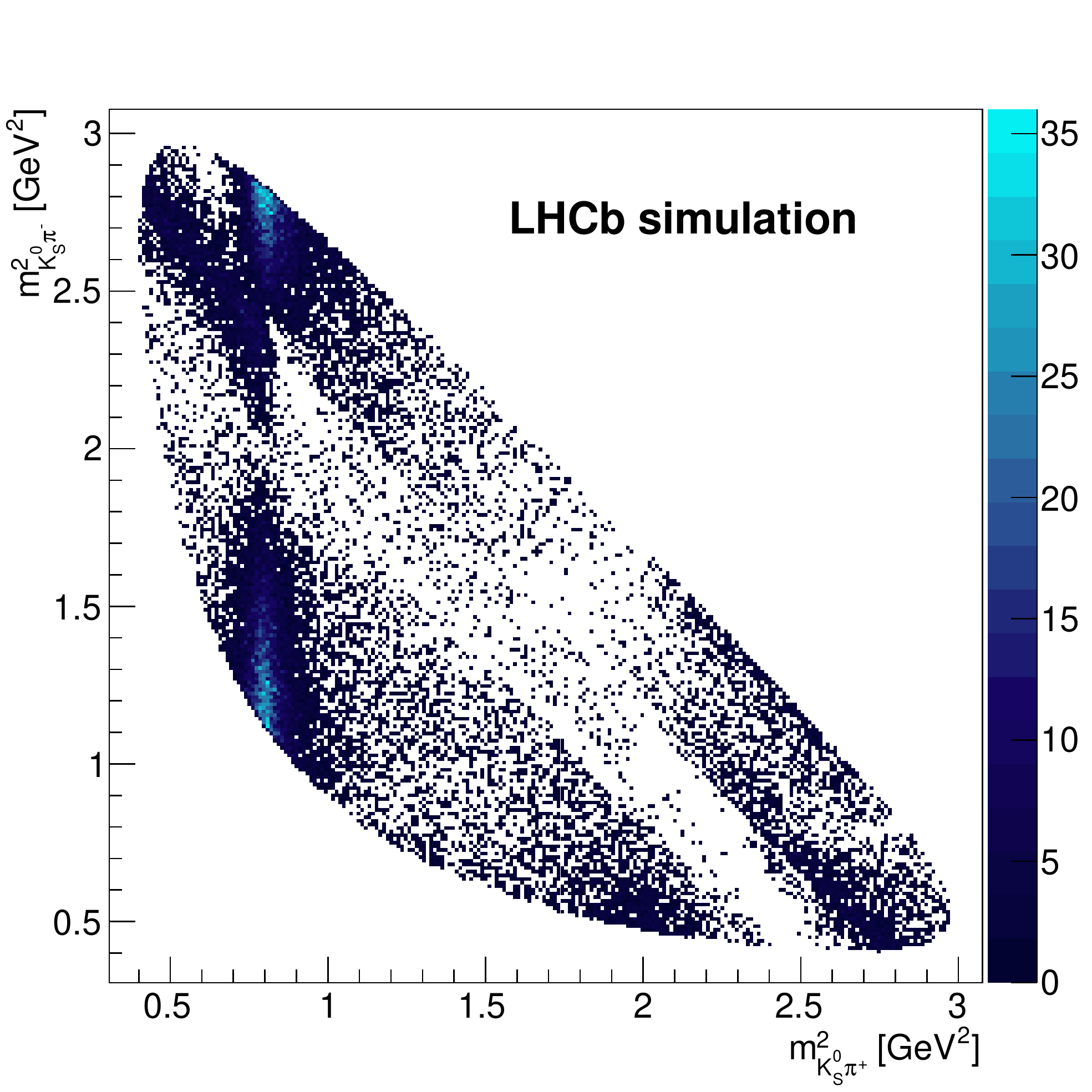}
\caption{Dalitz plot for \BDKshh decays produced from generator level Monte-Carlo simulation using the \babar 2010 model \cite{Babar2010}.}
\label{Dalitz_Toy}
\end{wrapfigure}
 
\DKshh decays can be formulated as quasi two-body decays via intermediate resonances where the parametrisation of the line shapes is model-dependent. The intermediate P- and D-wave resonances are the $\rho(770), f_2(1270), \omega(782)$ mesons for the $\pi^+\pi^-$ channel, the $K^*(892)^+, K^*_2(1430)^+$ mesons for the $K^0_S\pi^{+}$ mode and the  $K^*(892)^-, K^*_2(1430)^-, K^*(1680)^-$ mesons for the $K^0_S\pi^{-}$ contribution \cite{Babar2010}. With exception of the $\rho(770)$ meson which is described by a Gounaris-Sakurai distribution, the intermediate P- and D-wave resonances are modelled by a relativistic Breit-Wigner distribution \cite{Babar2010}. The S-wave contribution in the $K^0_S\pi^{\pm}$ channel corresponding to the $K^*_0(1430)^{\pm}$ meson is described by the LASS parametrisation whereas the K-matrix formalism is applied to formulate the $\pi^+\pi^-$ S-wave contributions \cite{Babar2010}. Changes to the model might be indispensable due to the achievable sensitivity. The interference of the listed amplitudes can be visualised in the Dalitz plane spanned by $m^2_{K^0_S\pi^+}$ and $m^2_{K^0_S\pi^-}$ as illustrated in Fig. \ref{Dalitz_Toy} for a sample of generator-level Monte-Carlo simulation sample of \BDKshh decays.\\

The time-dependent amplitude-model analysis fit will use GooFit \cite{GooFit} - a parallel fitting framework for Graphics Processing Units (GPUs) implemented in CUDA - which provides a significant speed-up compared to conventional Central Processing Units (CPUs). GooFit \cite{GooFit} is written especially for maximum-likelihood fits or time-dependent amlitude-model amplitude analyses and thus the most common line shape models like relativistic Breit-Wigner, Gounaris-Sakurai and the LASS parametrisation are available.

\section{Conclusion}

A measurement of the mixing parameters $x_D$ and $y_D$ as well as of the parameters $|q/p|$ and $\phi = arg(q,p)$, which govern indirect CP violation, will be performed based on a time-dependent amplitude-model analysis of the full LHCb dataset of 2011 and 2012 corresponding to an integrated luminosity of $3 \, \mathrm{fb^{-1}}$ using both prompt and semileptonically-tagged \DKshh decays.

\section*{Acknowledgements}

We express our gratitude to our colleagues in the CERN
accelerator departments for the excellent performance of the LHC. We
thank the technical and administrative staff at the LHCb
institutes. We acknowledge support from CERN and from the national
agencies: CAPES, CNPq, FAPERJ and FINEP (Brazil); NSFC (China);
CNRS/IN2P3 and Region Auvergne (France); BMBF, DFG, HGF and MPG
(Germany); SFI (Ireland); INFN (Italy); FOM and NWO (The Netherlands);
SCSR (Poland); MEN/IFA (Romania); MinES, Rosatom, RFBR and NRC
``Kurchatov Institute'' (Russia); MinECo, XuntaGal and GENCAT (Spain);
SNSF and SER (Switzerland); NAS Ukraine (Ukraine); STFC (United
Kingdom); NSF (USA). We also acknowledge the support received from the
ERC under FP7. The Tier1 computing centres are supported by IN2P3
(France), KIT and BMBF (Germany), INFN (Italy), NWO and SURF (The
Netherlands), PIC (Spain), GridPP (United Kingdom). We are thankful
for the computing resources put at our disposal by
Yandex LLC (Russia), as well as to the communities behind the multiple open
source software packages that we depend on.

\end{document}

%% file: charmmacros.tex
\newcommand\pubnumber{\pbnr}
\newcommand\pubdate{\today}
\def\Title#1{\begin{center} {\Large #1 } \end{center}}
\def\Author#1{\begin{center}{ \sc #1} \end{center}}

\newcommand{\OnBehalf}[1]{\sbox0{#1}\ifdim\wd0=0pt
        {}
	\else
	{\\on behalf of #1}
	\fi}
\newcommand{\SupportedBy}[1]{\sbox0{#1}\ifdim\wd0=0pt
        {}
	\else
	{\footnote{#1}}
	\fi}
\def\Address#1{\begin{center}{ \it #1} \end{center}}

\newcommand\pubblock{\includegraphics[width=5cm]{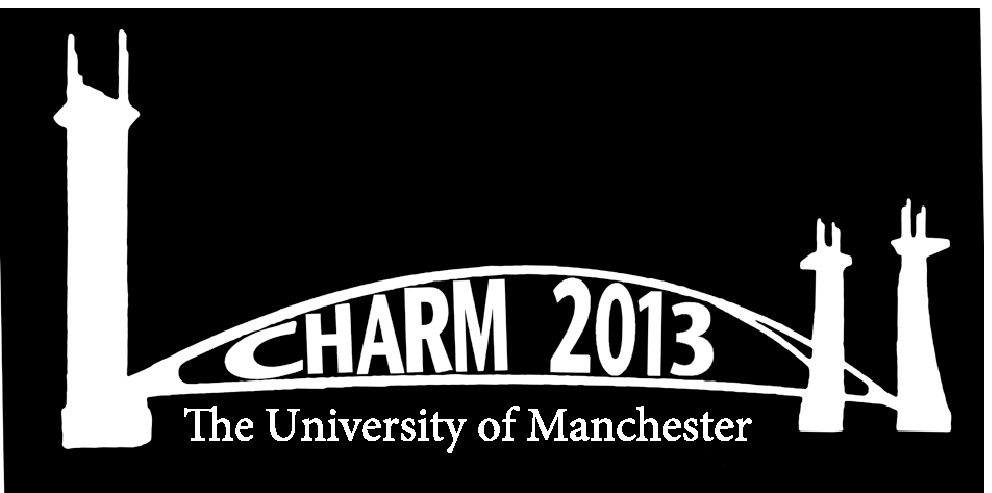}\hfill{\begin{tabular}{l} \pubnumber\\
         \pubdate  \end{tabular}}}
\newenvironment{Abstract}{\begin{quotation}  }{\end{quotation}}
\newenvironment{Presented}{\begin{quotation} \begin{center} 
             PRESENTED AT\end{center}\bigskip 
      \begin{center}\begin{large}}{\end{large}\end{center} \end{quotation}}

\def\venue{The 6$^{th}$ International Workshop on Charm Physics\\
(CHARM 2013)\\
Manchester, UK,  31 August -- 4 September, 2013}

\def\beq{\begin{equation}}
\def\eeq#1{\label{#1}\end{equation}}
\def\eeqn{\end{equation}}

\def\beqa{\begin{eqnarray}}
\def\eeqa#1{\label{#1}\end{eqnarray}}
\def\eeqan{\end{eqnarray}}

\let\bar=\overbar

\def\Dslash{\not{\hbox{\kern-4pt $D$}}}
\def\dslash{\not{\hbox{\kern-2pt $\del$}}}

\def\msb{{\bar{\ssstyle M \kern -1pt S}}}

